\documentclass[runningheads]{llncs}

\usepackage{multirow}
\usepackage{makeidx}  %for index
\usepackage{algorithm}
\usepackage[pdftex]{graphicx} %
\usepackage{amsmath,amssymb}
\usepackage[algoruled,linesnumbered,algo2e,vlined]{algorithm2e}
\usepackage{pdfpages}
\usepackage{array}
\usepackage{algorithmic}
\usepackage{setspace}
\usepackage{xspace}

\usepackage[hang,small,bf]{caption}
\usepackage[subrefformat=parens]{subcaption}
\usepackage[pagewise,mathlines]{lineno}
\let\doendproof\endproof
\renewcommand\endproof{~\hfill$\qed$\doendproof}
\usepackage[final,mode=multiuser]{fixme}

\fxsetup{theme=color}
\definecolor{fxtarget}{rgb}{0.0000,0.0000,0.4823}

\fxuseenvlayout{colorsig}
\fxusetargetlayout{color}

\fxuseenvlayout{colorsig}
\fxusetargetlayout{color}
\FXRegisterAuthor{yy}{ayy}{YY}
\FXRegisterAuthor{si}{asi}{SI}
\fxsetface{env}{}

\newcommand{\runs}[1]{\mathsf{Runs}(#1)}

\newcommand{\be}{\mathsf{beg}}
\newcommand{\fin}{\mathsf{end}}
\newcommand{\per}{\mathsf{p}}
\newcommand{\weight}{\mathsf{w}}

\mathchardef\mhyphen="2D

\newcommand{\Prefix}{\mathsf{Prefix}}
\newcommand{\Substr}{\mathsf{Substr}}
\newcommand{\Suffix}{\mathsf{Suffix}}

\newcommand{\SubRuns}{\mathsf{SubRuns}}

\newcommand{\bit}{\mathsf{b}}

\title{Simple Linear-time Repetition Factorization}
\author{
  Yuki~Yonemoto\inst{1}\orcidID{0009-0008-5330-7256} \and
  Shunsuke~Inenaga\inst{2}\orcidID{0000-0002-1833-010X} 
}
\institute{
  Department of Information Science and Technology, Kyushu University 
  \email{yonemoto.yuuki.240@s.kyushu-u.ac.jp} \and
  Department of Informatics, Kyushu University
  \email{inenaga.shunsuke.380@m.kyushu-u.ac.jp}
}

\begin{document}
\maketitle

\begin{abstract}
A factorization $f_1, \ldots, f_m$ of a string
$w$ of length $n$ is called a \emph{repetition factorization} of $w$
if $f_i$ is a repetition, i.e., $f_i$ is a form of $x^kx'$ where 
$x$ is a non-empty string, $x'$ is a (possibly-empty) proper prefix of $x$,
and $k \geq 2$. 
Dumitran et al. [SPIRE 2015] presented an $O(n)$-time and space algorithm for
computing an arbitrary repetition factorization of a given string of length $n$. 
Their algorithm heavily relies on 
the \emph{Union-Find} data structure on trees proposed by Gabow and Tarjan~[JCSS 1985] that works in linear time on the word RAM model,
and an interval stabbing data structure of Schmidt~[ISAAC 2009].
%However, from a practical point of view, this algorithm is complicated 
%because it uses bitwise operations on the word RAM model.
In this paper, we explore more combinatorial insights into the problem,
and present a simple algorithm to compute an arbitrary repetition factorization
of a given string of length $n$ in $O(n)$ time,
without relying on data structures for Union-Find and interval stabbing.
Our algorithm follows the approach by Inoue et al.~[ToCS 2022] that computes
the smallest/largest repetition factorization in $O(n \log n)$ time.

%As an alternative simpler solution, we propose a simple algorithm for computing an arbitrary repetition factorization in $O(n)$ time and space 
%that does not use Union-Find.

\end{abstract}

\section{Introduction}
A factorization of a string $w$ is a sequence 
$f_1, \ldots, f_k$ of non-empty strings of $w$ such that $w = f_1 \cdots f_k$.
The length $k$ of such a factorization is called the size of the factorization. 
String factorizations have widely been used for data compression~\cite{LZ77,LZ78}.
Also, factorizing a given string into combinatorial objects,
including palindromes~\cite{DBLP:journals/jda/FiciGKK14,DBLP:conf/cpm/ISIBT14,BorozdinKRS17,DBLP:journals/ijfcs/BannaiGIKKPS18}, squares~\cite{dumitran2015square,Matsuoka16}, repetitions~\cite{dumitran2015square,inoue2022factorizing,KishiNI23}, Lyndon words~\cite{Duval83}, and closed words~\cite{DBLP:journals/dam/BadkobehBGIIIPS16,AlzamelISS19}.
Among others, we focus on \emph{repetition factorizations} of strings first proposed by Dumitran et al.~\cite{dumitran2015square}, 
which has a close relation to the \emph{runs} (a.k.a. maximal repetitions)~\cite{DBLP:conf/focs/KolpakovK99,RunsTheorem,DBLP:conf/icalp/Ellert021}, 
which are one of the most significant combinatorial structures in stringology.

A period of a string $w$ of length $n$ is a positive integer $0 < p < n$ satisfying $w[i] = w[i+p]$ for any $1 \leq i \leq |w|-p$, 
and we represent the smallest period of a string $w$ as $\per(w)$.
Also, a substring $s$ of a string $w$ is called a repetition in $w$
if $2\per(s) \leq |s|$. 
Then, a factorization $f_1, \ldots, f_m$ of a string $w$
is called a repetition factorization of $w$ if each factor $f_i$ is a repetition. 
Dumitran et al.~\cite{dumitran2015square} presented an algorithm for computing an arbitrary repetition factorization of a string $w$ of length $n$ in $O(n)$ time and space.
Their algorithm heavily relies on 
the \emph{Union-Find} data structure on trees proposed by Gabow and Tarjan~\cite{GABOW1985209} that works in $O(N+m)$ time in the word RAM model,
where $N$ is the size of the tree and $m$ is the number of queries and operations.
In Dumitran et al.'s algorithm~\cite{dumitran2015square}
the tree is a list of length $n$ and $m = O(n)$,
and thus their algorithm runs in $O(n)$ time in the word RAM model.
Dumitran et al.~\cite{dumitran2015square} also use
the \emph{interval stabbing} data structure of Schmidt~\cite{schmidt2009interval} that is applied to the precomputed sets of runs.

%On the other hand,
%it is known that any Union-Find algorithm requires $\Omega(m \alpha(N+m,m)+N)$
%time on the pointer machine~\cite{Tarjan79}, which is a weaker model of computation.
%It is thus inferred that the precomputation of small tables
%and table lookups in the Union-Find data structure~\cite{GABOW1985209}
%are essential for Dumitran et al.'s repetition factorization algorithm to work in $O(n)$ time.

In this paper, we explore more combinatorial insights into the problem of finding an arbitrary repetition factorization,
and show that neither Union-Find structures nor interval stabbing structures are necessary.
Our resulting algorithm is quite simple and works in $O(n)$ time and space. 
Our algorithm is based on the idea from Inoue et al.~\cite{inoue2022factorizing}
that computes the repetition factorizations of the smallest/largest size
in $O(n \log{n})$ time.
Their algorithm builds a \emph{repetition graph} over an input string $w$,
that is a weighted directed graph representing all repetitions in $w$.
A shortest (resp. longest) source-to-sink path of the repetition graph corresponds to a smallest (resp. largest) repetition factorization of $w$.
The bottleneck is that the number of nodes and edges of the repetition graph is $\Omega(n \log n)$~\cite{inoue2022factorizing} in the worst case.
In this paper, we propose a sparse version of the graph, called the \emph{arbitrary repetition factorization graph} (\emph{ARF} graph), of size $O(n)$ for any string of length $n$,
which is a key data structure for our $O(n)$-time solution.

The rest of this paper is organized as follows:
In section~\ref{ch:preliminary}, we present preliminaries of this paper.
Especially, in Section~\ref{sec:pre_rep}, we present preliminaries of repetitive structures in strings, and define the problem considered in this paper.
In section~\ref{ch:rep}, we describe our method.
Especially, in Section~\ref{sec:rep_idea}, we recall previous work and present the main idea of our algorithm.
Also, in Section~\ref{sec:rep_alg}, we present our algorithm and prove the following:

\begin{theorem} \label{th:rep}
  Given a string $w$ of length $n$, we can compute an arbitrary repetition factorization 
  of $w$ in $O(n)$ time and $O(n)$ space without bit operations in the word RAM model. 
\end{theorem}

%presented algorithms for computing smallest/largest size repetition factorizations. 
%Since our algorithm has a close relation to their algorithms, 
%we recall the algorithms of Inoue et al. in Section~\ref{ch:rep} for clarity of our algorithm.

%Their algorithm for repetition factorizations includes the algorithm 
%that computes the \emph{Union-find} structure in linear-time under special cases~\cite{GABOW1985209}. 
%However, from a practical point of view, this algorithm is complicated 
%because it uses bitwise operations in the word RAM model,
%also, from an application point of view, this algorithm is complicated because 
%it requires a precise analysis of the precondition, such that we need prior knowledge of the overview of the Union-find data structure.

%As an alternative simpler solution, we propose a simple algorithm for computing an arbitrary repetition factorization in $O(n)$ time and space 
%that does not use Union-find.

\section{Preliminary} \label{ch:preliminary}

\subsection{Strings}
Let $\Sigma$ be an alphabet.
An element of $\Sigma^*$ is called a \emph{string}.
The \emph{length} of a string $w$ is denoted by $|w|$. 
The \emph{empty string}, denoted by $\varepsilon$, 
is a string of length $0$.
Let $\Sigma^+ = \Sigma^\ast \setminus \{\varepsilon\}$.
For a string $w = xyz$ with $x,y,z \in \Sigma^*$,
strings $x$, $y$, and $z$ are called a \emph{prefix}, \emph{substring}, 
and \emph{suffix} of string $w$, respectively.
For a string $w$, let $\Prefix(w)$, $\Substr(w)$ and $\Suffix(w)$ denote the sets of
prefixes, substrings and suffixes of $w$, respectively.
The strings in $\Prefix(w) \setminus \{w\}$ (resp. $\Suffix(w) \setminus \{w\}$)
are called \emph{proper prefixes} (resp. \emph{proper suffixes}) of $w$. 
The $i$-th character of a string $w$ is denoted by $w[i]$ for $1\le i\le|w|$, and
the substring of $w$ that begins at position $i$ and ends at position $j$
is denoted by $w[i..j]$ for $1\le i\le j\le |w|$.
For convenience, let $w[i..j]=\varepsilon$ for $i>j$.
%
%A string $u$ is a \emph{subsequence} of another string $w$
%if $u = \varepsilon$ or there exists a sequence of integers $i_1,\ldots,i_{|u|}$ 
%such that $1\le i_1<\cdots<i_{|u|}\le|w|$ and $u=w[i_1]\cdots w[i_{|u|}]$.

\subsection{Repetitive structures} \label{sec:pre_rep}
%We define the repetitive structures used in this paper.

A \emph{period} of a string $w$ is a positive integer $0 < p < n$ satisfying $w[i] = w[i+p]$ for any $1 \leq i \leq |w|-p$.
Let $\per(w)$ denote the \emph{smallest} period of a string $w$.
A string $w$ is called \emph{primitive} if $w$ cannot be expressed as $w=x^k$
with any string $x$ and integer $k \geq 2$.
A non-empty string $w$ is called a \emph{square} if $w=x^2$ for some string $x$.
A square $x^2$ is called a \emph{primitively rooted square} if $x$ is primitive.  
A substring $s$ of another string $w$ is called a \emph{repetition} in $w$
if $2\per(s) \leq |s|$.
Alternatively, a repetition $s$ has a form $s = x^kx'$ where
$k \geq 2$ is an integer and $x'$ is a proper prefix of $x$.
%For example, $s = \mathtt{baabaaba}$ is a repetition with period $3$.
In what follows, we consider an arbitrarily fixed string $w$,
and we represent a repetition $s = w[\be..\fin]$ in $w$ by
a tuple $(\be, \fin, \per(s))$.

A \emph{run} in a string $w$ is a repetition $s = w[\be..\fin]$ satisfying the following two conditions:
\begin{itemize}
  \item $\be = 1$ or $w[\be - 1] \neq w[\be + \per(s) - 1]$ and,
  \item $\fin = |w|$ or $w[\fin + 1] \neq w[\fin - \per(s) + 1]$.
\end{itemize}

Let $\runs{w}$ denote the set of the runs in a string $w$,
and let $|\runs{w}|$ denote the number of runs in $\runs{w}$.
Kolpakov and Kucherov~\cite{DBLP:conf/focs/KolpakovK99} showed that $|\runs{w}| = O(n)$ for any string $w$ of length $n$,
and later Bannai et al.~\cite{RunsTheorem} proved that $|\runs{w}| < n$.
Ellert and Fischer~\cite{DBLP:conf/icalp/Ellert021} showed how to compute $\runs{w}$ for a given string $w$ of length $n$ over an general ordered alphabet in $O(n)$ time.
We can then sort $\runs{w}$ based on their beginning positions of runs in non-decreasing order in linear time by bucket sort.
Let $r_i=(\be_i, \fin_i, \per_i)$ denote the $i$-th run in $\runs{w}$ sorted in increasing order of $\be_i$, and of $\fin_j-2\per_j$ if draw.

\subsection{Repetition factorizations}

For a given string $w$ of length $n$, a sequence $f_1, \ldots, f_m$ of non-empty strings 
is said to be a \emph{repetition factorization} of $w$
if the sequence satisfies the following two conditions:
\begin{itemize}
  \item $w =  f_1 \cdots f_m$, and 
  \item each $f_i$ $(1 \leq i \leq n)$ is a repetition in $w$.
\end{itemize}
Each $f_i$ is called a \emph{factor}, and $m$ is called the size of the factorization of $w$.
For example, string $w = \mathtt{abaababababaabaab}$ has 
a repetition factorization $f_1, f_2, f_3$ with $f_1 = \mathtt{abaaba}$, $f_2 = \mathtt{baba}$, and $f_3 = \mathtt{baabaab}$.
Note also that some strings do not have a repetition factorization.

In this paper, we tackle the following problem:
\begin{problem}[Computing an arbitrary repetition factorization]\label{prob:rep_f}
  \leavevmode
  \begin{description}
    \item[Input:] A string $w$ of length $n$.
    \item[Output:] An arbitrary repetition factorization of $w$ if $w$ has a repetition factorization,
    and ``no'' otherwise.
  \end{description}
\end{problem}

For a reference, we introduce the problem considered by Inoue et al.~\cite{inoue2022factorizing}:
\begin{problem}[Computing smallest/largest repetition factorization]\label{prob:rep_f_sl}
  \leavevmode
  \begin{description}
    \item[Input:] A string $w$ of length $n$.
    \item[Output:] A repetition factorization of smallest/largest size if $w$ has a repetition factorization,
    and ``no'' otherwise.
  \end{description}
\end{problem} 

For example, for string $w = \mathtt{aabaabaacbbcbbcbb}$, 
a smallest repetition factorization of $w$ is $f_1, f_2$  of size $2$ with 
$f_1 = \mathtt{aabaabaa}$ and $f_2 = \mathtt{cbbcbbcbb}$,
and a largest repetition factorization of $w$ is $g_1, g_2, g_3, g_4$ of size $4$ with
$g_1 = \mathtt{aabaab}$, $g_2 = \mathtt{aa}$, $g_3 = \mathtt{cbbcbbc}$, and $g_4 = \mathtt{bb}$.

\section{Simple linear-time repetition factorization algorithm} \label{ch:rep}

In this section, we present a simple linear-time algorithm for computing an arbitrary repetition factorization of a given string, which does not use bit operations in the word RAM model.

\subsection{Overview of our approach} \label{sec:rep_idea}
In this subsection, we describe the basic ideas of our algorithm, following the notations from Inoue et al.~\cite{inoue2022factorizing}:
For an input string $w$ of length $n$,
Inoue et al. introduced the repetition graph $G_w$ of $w$ of size $O(n \log n)$,
such that the smallest/largest repetition factorization problem
is reducible to the shortest/longest path problem on $G_w$.
This solves Problem~\ref{prob:rep_f_sl} in $O(n \log n)$ time.

To solve Problem~\ref{prob:rep_f} in linear time,
we introduce the \emph{arbitrary repetition factorization graph} (\emph{ARF-graph})
which is a compact version of the repetition graph for finding an arbitrary repetition factorization.
While the ARF-graph is still of size $O(n \log n)$,
it permits us to compute a solution to Problem~\ref{prob:rep_f}
in $O(n)$ time without explicitly constructing the whole graph.

\subsection{Repetition graphs} \label{subsec:graph_def}

\begin{figure}[tbh]
  \centerline{
    \includegraphics[width = 1.0\textwidth]{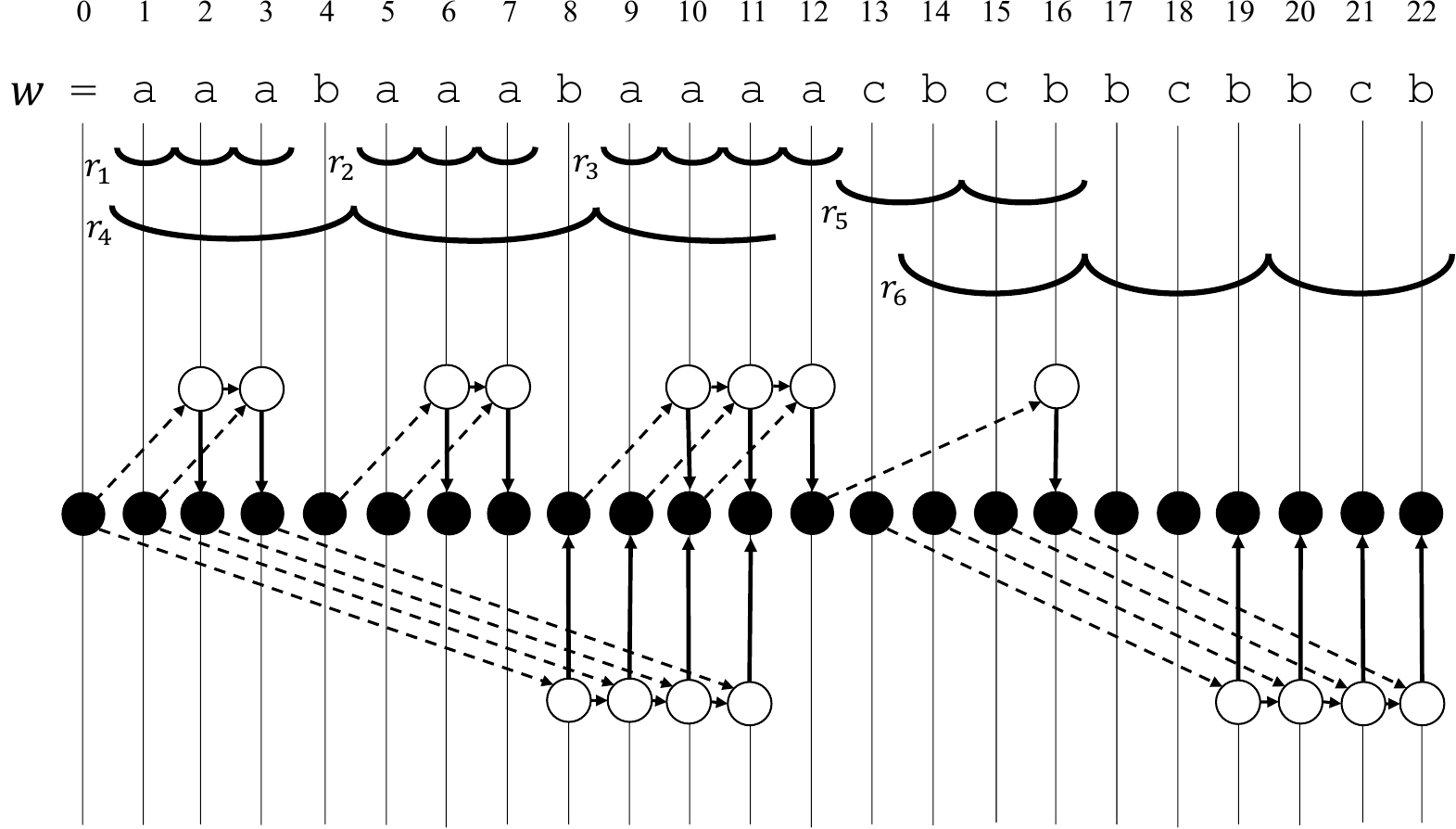}
  }
  \caption{
    This figure shows the repetition graph of a string $w=\mathtt{aaabaaabaaaacbcb}$ $\mathtt{bcbbcb}$.
    The concatenations of curves at the top represent each run of $w$. 
    The nodes in $V'$ are illustrated as black circles, whose positions correspond to each character of $w$,
    the nodes in $V''$ are illustrated as white circles for each run,
    the edges in $E'$ are illustrated as dashed arrows,
    and the edges in $E''$ and $E'''$ are illustrated as horizontal and vertical solid arrows, respectively.
  }
  \label{fig:rep_graph}
\end{figure}

Let us recall the method of Inoue et al.~\cite{inoue2022factorizing}.
The repetition graph $G_w = (V, E)$ of a string $w$ is the following
directed acyclic edge-weighted graph with weight function $\weight: V \rightarrow \mathcal{N}$.
The set $V = V' \cup V''$ of nodes consists of two disjoint sets of nodes $V'$ and $V''$ such that
\begin{eqnarray*}
  V' & = & \{ (0, j) \mid 0 \leq j \leq n \},\\
  V'' & = & \bigcup_{i=1}^{|\runs{w}|} V''_i,
\end{eqnarray*}
where
$V''_i = \{ (i, j) \mid \be_i + 2\per_i - 1 \leq j \leq \fin_i \}$
for each $r_i = (\be_i, \fin_i, \per_i) \in \runs{w}$.
Intuitively, the subset $V'$ is the nodes representing all $n+1$ positions in $w$, including $0$ that is the source of $G_w$.
We call these nodes as \emph{black nodes} (see also Fig.~\ref{fig:rep_graph}).
The other subset $V''$ of nodes represents the positions included in
the runs from $\runs{w}$, but excluding the first $2p-1$ positions
for each run $r = x^kx'$ with $|x| = p$ (the white nodes in Fig.~\ref{fig:rep_graph}).
We call these nodes as \emph{white nodes}.
For each node $v = (i,j) \in V$,
$j$ is said to be the \emph{position} of the node $v$
and we denote it by $p(v) = j$.

The set $E = E' \cup E'' \cup E'''$ of edges consists of three disjoint sets of edges $E'$, $E''$, and $E'''$ such that
\begin{eqnarray*}
  E' & = & \{ ((i_1, j_1),(i_2, j_2)) \mid j_2 - j_1 = 2\per_{i_2}, (i_1, j_1) \in V', (i_2, j_2) \in V'' \}, \\
  E'' & = & \{ ((i_1, j_1),(i_2, j_2)) \mid i_1 = i_2, j_1 + 1 = j_2, (i_1, j_1), (i_2, j_2) \in V'' \}, \\
  E''' & = & \{ ((i_1, j_1),(i_2, j_2)) \mid j_1 = j_2, (i_1, j_1) \in V'', (i_2, j_2) \in V' \}.
\end{eqnarray*}
For each edge $e \in E$, $\weight(e)$ is defined as follows:
\[
  \weight(e) =
  \begin{cases}
    1 & \mbox{if } e \in E', \\
    0 & \mbox{otherwise}.
  \end{cases}
\]

Intuitively, the nodes and edges of $G_w$ have the following structure:
\begin{itemize}
\item Each black node $(0, j) \in V'$ represents the position $j$ of each character in $w$.
\item Each white node $(i, k) \in V''$ represents the end position $k$ of a repetition contained in the $i$th run $r_i$.
\item Each edge $(u, v) \in E'$ with black node $u = (0,j-1) \in V'$ and white node $v = (i, k) \in V''$
represents a square of period $(k-j+1)/2$
that begins at position $j$ and ends at position $k$ in the $i$th run $r_i$.
By setting the origin node of each edge in $E'$ to the immediately preceeding position $j-1$, any two consecutive repetitions from different runs are connected and form a path in $G_w$.
\item Each edge in $E''$ connecting white nodes represents an operation of extending
a repetition to the right by one character within the corresponding run.
\item Each edge in $(u, v) \in E'''$ with white node $u = (i,j) \in V''$ and black node $v = (0,j) \in V'$ represents a (possible) boundary position $i$ of a repetition factorization in $w$.
\end{itemize}

Fig.~\ref{fig:rep_graph} shows an example of a repetition graph.
For example, the repetition $w[14..20] = \mathtt{bcbbcbb}$ in $w$ is represented 
by the path from the black node $(0,13)$ to the black node $(0,20)$ which goes through
the diagonal edge from $(0,13)$ to the white node $(6, 19)$, 
the horizontal edge from $(6,19)$ to $(6,20)$, and 
the vertical edge from $(6,20)$ to $(0,20)$.

\sinote*{reworded}{%
Clearly $|V'| = n$. 
Inoue et al.~\cite{inoue2022factorizing} showed that 
$|V''| = O(n\log{n})$ using Lemma~\ref{lem:prsq} below.
Since $|E| = O(|V|)$,
the size of the repetition graph is $O(n\log{n})$.
}%

\begin{lemma}[\cite{CrochemoreR95}] \label{lem:prsq}
  The number of primitively rooted squares which appear in suffixes of $w[1..i]$ is $O(\log{i})$ for a string $w$. 
\end{lemma}

For convenience, we will denote $(0, 0) \in V'$ by $v_s$.
Inoue et al.~\cite{inoue2022factorizing} showed a correspondence between a path on $G_w$ and a repetition factorization of $w$ by proving Lemma~\ref{lem:correspondence_lemma} shown below.

\begin{lemma}[\cite{inoue2022factorizing}] \label{lem:correspondence_lemma}
  For any string $w$ of length $n$ and integer $1 \leq t \leq n$,
  repetition factorizations of $w[1..t]$ and $s \mhyphen t$ paths
  have a one-to-one correspondence,
  where a $s \mhyphen t$ path is a path from $v_s$ to $(0,t) \in V'$ in $G_w$.
\end{lemma}

From Lemma~\ref{lem:correspondence_lemma}, we can compute the smallest/largest repetition factorizations 
in $O(n\log{n})$ time and space based on dynamic programming and backtracking on the repetition graph,
such that each node stores the maximum/minimum value of the weight of a path starting at $v_s$.

Also, we can compute an arbitrary repetition factorization 
in $O(n\log{n})$ time and space based on dynamic programming and backtracking on the repetition graph,
such that each node stores the $0/1$ value depending on whether it is unreachable/reachable from $v_s$.

\subsection{ARF-graphs} \label{subsec:graph_def_2}

\begin{figure}[tb]
  \centering
  \begin{minipage}[b]{1\columnwidth}
  \centering
  \includegraphics[scale=0.35]{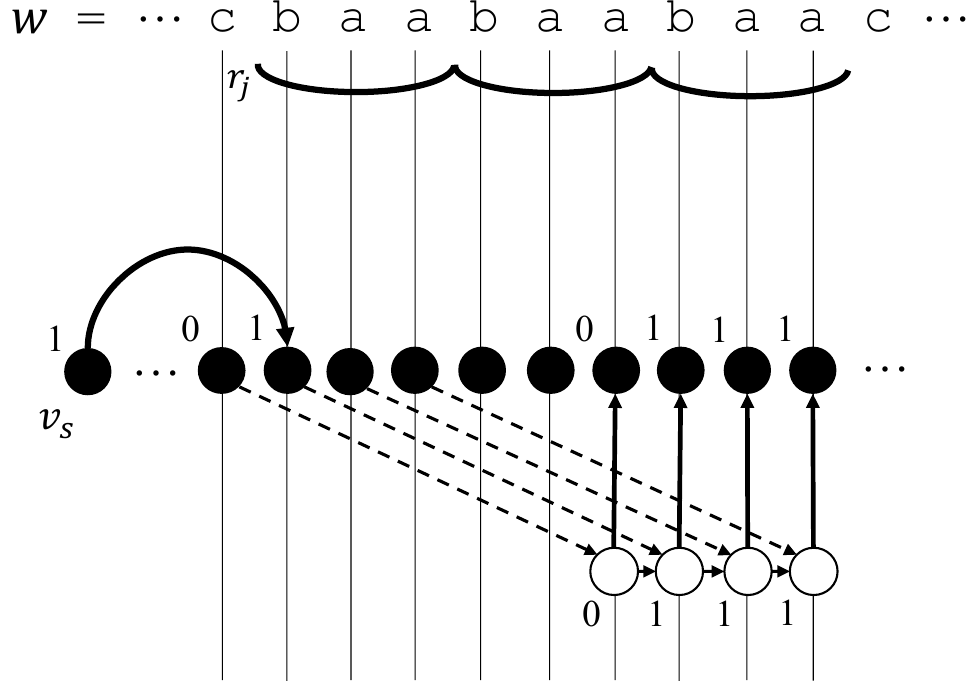}
  \subcaption{
  }
  \label{fig:rep_1}
  \vspace{2mm}
  \hspace{-1cm}
  \begin{minipage}[b]{0.4\columnwidth}
    \centering
    \includegraphics[scale=0.35]{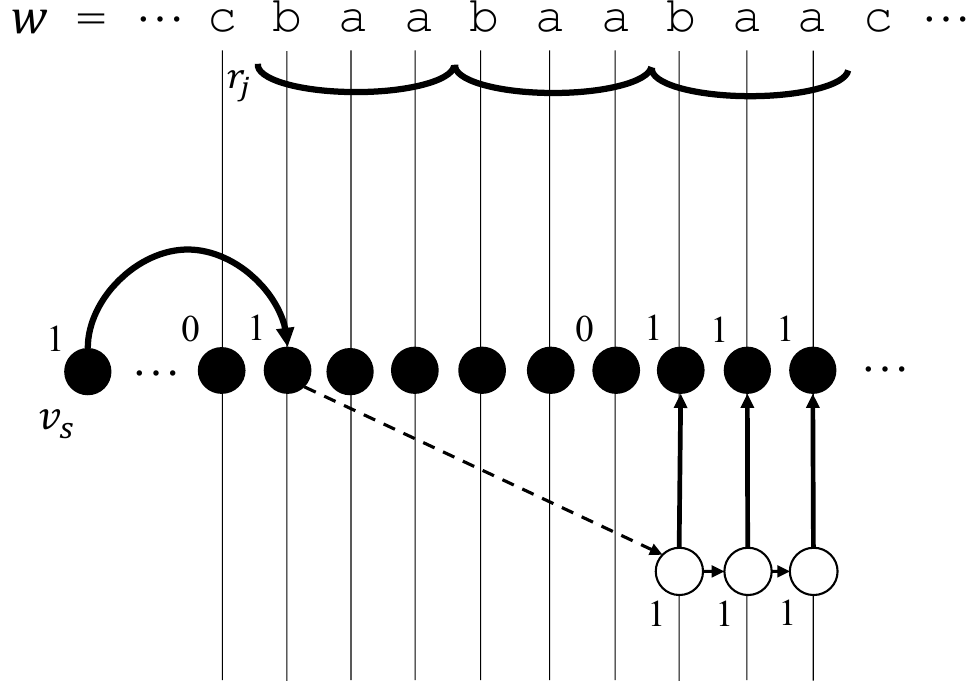}
    \subcaption{
    }
    \label{fig:rep_2}
  \end{minipage}
  \hspace{0.1\columnwidth}
  \begin{minipage}[b]{0.4\columnwidth}
    \centering
    \includegraphics[scale=0.35]{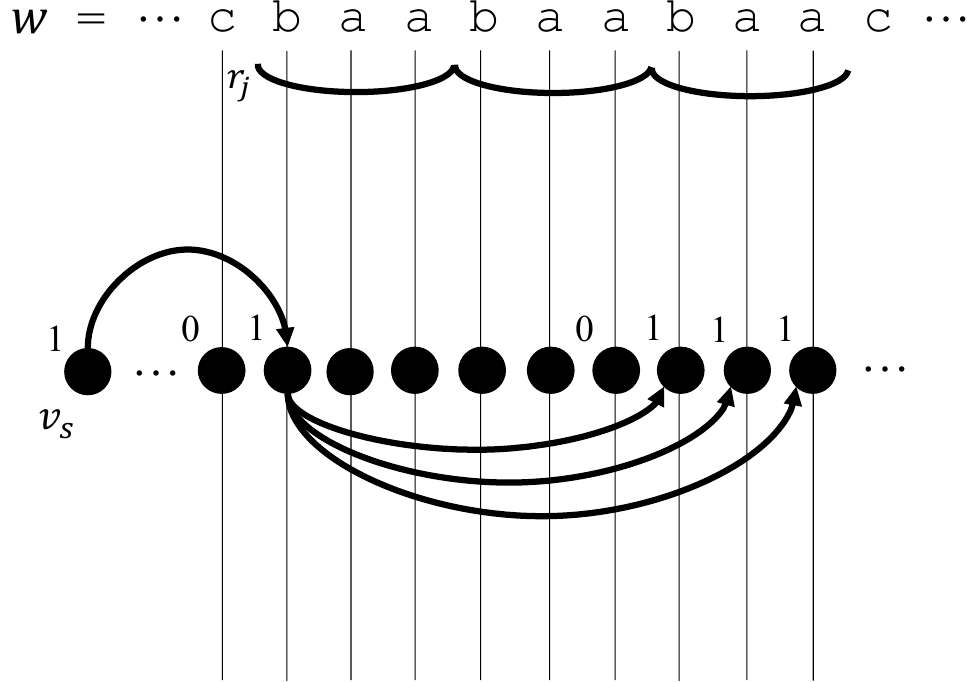}
    \subcaption{
    }
    \label{fig:rep_3}
  \end{minipage}
  \end{minipage}
  \caption{
    Parts of repetition graph $G_w$ (Fig.~\ref{fig:rep_1}),
    intermediate graph $H_w$ (Fig.~\ref{fig:rep_2}),
    and ARF-graph $\hat{G}_w$ (Fig.~\ref{fig:rep_3}).
  }
%  Fig.~\ref{fig:rep_1} shows the part of the repetition graph of 
%  $w = \cdots \mathtt{cbaabaabaac} \cdots$ which has a run $r_j$ of $\mathtt{baabaabaa}$.
%  Fig.~\ref{fig:rep_1} shows the graph excluding edges connected to or from the white node $\hat{v}_7$ corresponding to $w'[7]$,
%  and the edges in $E'$ connected to white nodes $\hat{v}_9$ and $\hat{v}_{10}$ corresponding to $w'[9]$ and $w'[10]$,
%  from Fig.~\ref{fig:rep_1}, where $w'$ is the substring $\mathtt{cbaabaabaa}$ of $w$. 
%  Fig.~\ref{fig:rep_3} shows the graph omitted the white nodes and some edges connected to and from these nodes,
%  from Fig.~\ref{fig:rep_2}.}
%  \label{fig:rep_1-3}
\end{figure}

When it comes to computing an \emph{arbitrary} repetition factorization,
there is some redundant information in the repetition graph.
See Fig.~\ref{fig:rep_1}, where we consider a part of the repetition graph of 
a string $w$ that has substring $\mathtt{cbaabaabaac}$, in which there is a run $r_j$ $\mathtt{baabaabaa}$.
Let us assign a one bit information $\bit$ to each node $v$ of the repetition graph $G_w$,
such that $\bit(v) = 1$ iff $v$ is reachable from the source $v_s$.
For each subset $V_i''$ of white nodes corresponding to a run $r_i$,
\sinote*{changed}{%
we focus on the leftmost white node $w_i$ with $\bit(w_i) = 1$.
It is clear that any nodes $y$ in $V_i''$ to the left of $w_i$ are redundant (unreachable from $v_s$), and
any other nodes $z$ to the right of $w_i$ in $V_i''$ are reachable from $v_s$.
Thus we can remove all such nodes $y$.
Also, since there is an alternative path from $v_s$ to each $z$ via $u$,
we can also remove the diagonal edge to $z$.
This gives us an intermediate graph $H_w$ shown in Fig.~\ref{fig:rep_2}.
Further, we represent the path that begins with
the diagonal edge $(u_i, w_i)$ and arrives at a black node $x$
with a single edge from the origin black node $u_i$ to the target black node $x$.
Note that node $x$ remains reachable from the source $v_s$.
The resulting graph is the ARF-graph $\hat{G}_w$, illustarted in Fig.~\ref{fig:rep_3}.
}%

\begin{figure}[tb]
  \centerline{
    \includegraphics[width = 1.0\textwidth]{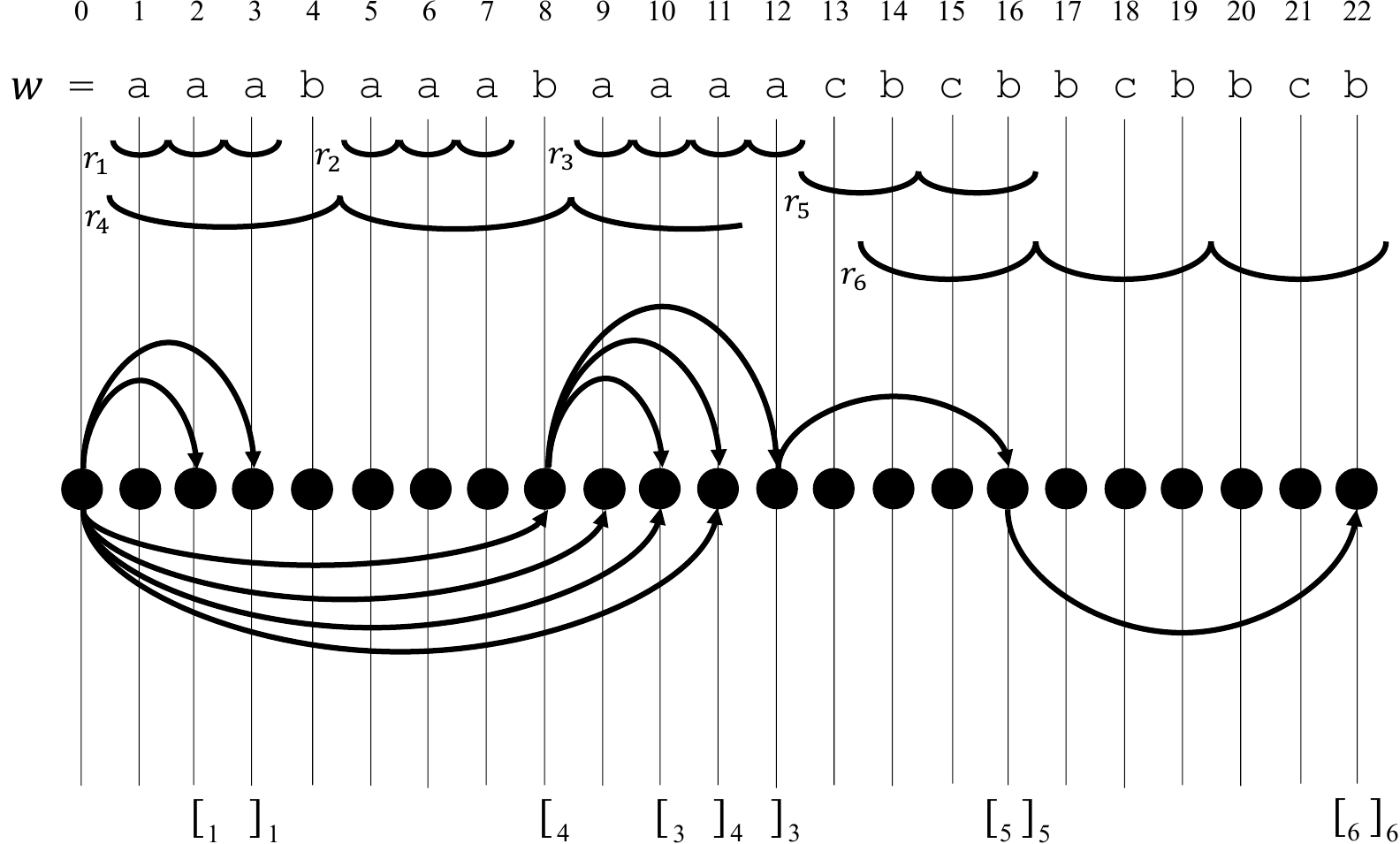}
  }
  \caption{
    The figure shows the ARF-graph
    factorization of a string $\mathtt{aaabaaabaaaa}$ $\mathtt{cbcbbcbbcb}$.
    The square brackets at the bottom represent the intervals for the consecutive incoming edges.
  }
  \label{fig:rep_graph_new}
\end{figure}

Formally, the ARF-graph $\hat{G}_w = (\hat{V}, \hat{E})$ for a string $w$ is given by
\sinote*{modified}{%
\begin{eqnarray*}
\hat{V} & = & \{ v_j \mid 0 \leq j \leq n \}, \\
\hat{E} & = & \bigcup_{i = 1}^{|\runs{w}|} \hat{E_i} \\
  & = & \bigcup_{i = 1}^{|\runs{w}|}\{ (u_i, x) \mid p(u_i) + 2\per_i \leq p(x) \leq \fin_i \},
%  & = & \bigcup_{i = 1}^{|\runs{w}|}\{ (u, v) \mid \be_i-1 \leq p(u) \leq \fin_i-2\per, \;\; p(v) - p(u) \geq 2\per_i \},
\end{eqnarray*}
}%
where, for each run $r_i = (\be_i, \fin_i, \per_i)$,
$\hat{E}_i$ is the set of edges from 
the leftmost black node $u_i$ in the position range $[\be_i-1,\fin_i-2\per_i]$
reachable from the source (i.e. $\bit(u_i) = 1$),
to the target black nodes $x$ within the run $r_i$ that are at least $2\per_i$ aray from $u_i$.
Fig.~\ref{fig:rep_graph_new} illustrates a concrete example of the ARF-graph.

%The set $\hat{E}$ of edges is defined as:
%\begin{equation*}
%  \hat{E} = \bigcup_{j=1}^{|\runs{w}|} 
%  \left\{
%    (v_{i_1}, v_{i_2}) ~\middle|~
%    \begin{array}{l}
%      v_{i_1} \mbox{satisfies that $i_1$ is the smallest index of the } \\
%      \mbox{starting node of edges in $\hat{E}_j$}, \;\; (v_{i_1}, v_{i_2}) \in \hat{E}_j
%    \end{array}
%  \right\}
%\end{equation*}
%where $\hat{E}_j$ is the set of edges $(v_{i_1}, v_{i_2})$ satisfying that 
%$v_{i_1}$ can be reached from $v_s$ in the graph 

\subsection{How to exclude redundant edges from the ARF-graph} \label{subsec:kakko}

The size of the ARF-graph remains $O(n\log{n})$
since the number of edges remains $O(n\log{n})$.
Therefore, we cannot afford to explicitly constructing all the edges of
the ARF-graph $\hat{G}_w$ for a given string $w$ of length $n$.
Still, our new method allows one to compute a solution in linear time. 

Before describing our new algorithm,
we revisit the previous method by Dumitran et al.~\cite{dumitran2015square}
and recall their algorithm in terms of the ARF-graph.
We emphasize that the ARF-graph was not present in their paper~\cite{dumitran2015square}.

\subsubsection{Algorithm of Dumitran et al.}
The algorithm of Dumitran et al.~\cite{dumitran2015square} 
computes a solution (i.e. an arbitrary repetition factorization of $w$) in linear time,
with a help of the interval Union-find data structure~\cite{GABOW1985209}
and without explicitly constructing all the edges of the ARF-graph.
For the sake of clarity, we represent each element in the Union-Find structure is an interval $[a,b]$ $(1 \leq a \leq b \leq n)$.
Initially, there are exactly $n$ intervals of form $[i,i]$ for all $1 \leq i \leq n$.
One executes the following two basic operations and an additional operation in this structure. 
One is the union operation for two intervals, 
and the other is the find operation which returns the start and end positions of the interval containing a given position on the ARF-graph.
In Gabow and Tarjan's algorithm~\cite{GABOW1985209}, the time complexity of executing these operations is $O(n+m)$ for $n$ intervals and $m$ operations. 
The additional one is marking operations for each interval of the above structure, where all the intervals are initially unmarked.
These operations are done so that later one can determine whether the value of each node has already been updated or not (corresponding to being marked or unmarked, respectively).  

%\sinote*{added}{%
%For convenience, we use a convention that
%the ARF-graph has $n+1$ positions from $0$ to $n$,
%and that each node $(i,j)$ is at position $j$ in the graph.
%}%

Initially, set $\bit(v_s) \leftarrow 1$ and $\bit(v) \leftarrow 0$ for all nodes $v$ of $\hat{G}_w$ except for $v_s$.
The one bit information $\bit(v)$ is then updated to $1$
iff $v$ is reachable from the source, in $O(n)$ total time,
with the Union-Find structure as follows:

%\sinote*{their algorithm?}{In our algorithm},
%one achieves $O(n)$-time computation by updating the
%one bit information of each node of the ARF-graph.

Initially set $R = \runs{w}$.
The algorithm executes the following queries and operations for node $v_j$
\yynote*{change}{%
  at each position $j$ with $\bit(v_j) = 1$ for $1 \leq j \leq n$ in increasing order:
}%
\begin{itemize}
  \item Retrieve all runs $r_i=(\be_i, \fin_i, \per_i)$ satisfying $\be_i-1 \leq j \leq \fin_i-2\per_i+1$ from $R$, 
  and execute the following to each computed run $r_i$:
  \begin{itemize}
    \item For the interval $[s,t] = [j+2\per_i,\fin_i]$ (this interval is not an interval in the Union-find structure) of the positions of the nodes that can be reached from $v_j$ by the edges of $r_i$, 
      execute a find query for the position $s$.
      Let $[c, c']$ be the obtained interval.
    \item If $[c,c']$ is unmarked, mark the interval and update the value of the node of position $t$ to $1$. Execute this operation for all the intervals obtained by the find queries.
    \item If $[c,c']$ does not contain $t$, execute a find query for given position $c'+1$, and let $[c'+1, c'']$ be the obtained interval.
    \item Execute a union operation to $[c,c']$ and $[c'+1,c'']$.
    \item If the new interval $[c, c'']$ does not contain $t$, 
    execute the union operation for the new interval $[c,c'']$ and the interval $[c'', c''']$ containing $c''$. 
    Continue executing this series of operations until one obtains the interval containing $t$. 
    \item Remove the run $r_i$ from $R$.
  \end{itemize}
\end{itemize}
After this computation, all values of the nodes of positions between $s$ and $t$ are $1$.
\yynote*{change}{%
Since the number of the union operations for each position is at most one, 
}%
the total time complexity for the union operations is $O(n)$.
Also, for all computations for all runs and positions, the number of times to update the values of nodes is $O(n)$.

In what follows, we present another idea for computing
an arbitrary repetition factorization of a given string
based on the ARF-graph.

\subsection{New Algorithm} \label{sec:rep_alg}

\subsubsection{Excluding Union-Find.}
The ending nodes of the edges that correspond to the same run are continuous in the ARF-graph,
and we decide whether each node can be reached from $v_s$ or not using these edges.
Therefore, the edges for a run which can be reached from $v_s$ can be expressed as an interval. 
The start point of the interval is the left-maximal node which has in-coming edges in $V_r$,
and the end point of the interval is the right-maximal node which has in-coming edges in $V_r$,
where $V_r$ is the set of nodes of the ARF-graph for run $r$.
Fig.~\ref{fig:rep_graph_new} shows examples of interval expressions for edges as square brackets.
For example, the interval $[8..11]$ expresses the edges of the run $r_4$.
Thus, the node can be reached from $v_s$, iff the position of that node is in any intervals.
Therefore, we can compute the value of each node by counting the number of the start points and end points of intervals by the following method. 
For each position $i$, we first compute whether $i$ is in any intervals or not, and if it is, we update the value of the node of position $i$ to $1$. 
We can compute the conditions dynamically by counting the number of the start/end points of intervals from the position $0$ to $i$.
Then, we execute the following operations for each position $i$ of the ARF-graph such that the value of the node of position $i$ is $1$: 
we compute the list of the runs $r_j=(\be_j, \fin_j, \per_j)$ satisfying $\be_j-1 \leq i \leq \fin_j-2\per_j+1$, 
and execute the following operations for each run $r_j=(\be_j, \fin_j, \per_j)$ in the list:
we mark the start and end point of the interval $[i+2\per_j,\fin_j]$ to nodes of positions $i+2\per_j$ and $\fin_j$, and remove $r_j$.

\subsubsection{Algorithm.} % \label{sec:rep_alg}
Our algorithm has two steps which are the step of deciding solution existence and the step of reconstructing a solution.  
First, we present the step for deciding whether there is a solution.

From the above idea, our algorithm for the step of deciding solution existence is shown in Algorithm~\ref{alg:decide}.
In this algorithm, the value of each node is stored in an array $M$ whose index begins with $0$, 
and the number of the start/end points of intervals for each node is stored in an array $L$/$R$,
and an array $P$ stores the start node of edges in the ARF-graph for each run, 
\yynote*{add description of $P$}{%
  which is used to obtain the solution in the latter step.
}%
If there are no edges for run $r_j$, $P[j] = -1$ (the elements of $P$ are initialized $-1$). 
$C$ is a counter for deciding that each node is in any interval.
$L$ and $R$ are updated dynamically from $M$, namely, each interval is computed dynamically.
In the lines $5$ and $14$ in Algorithm~\ref{alg:decide}, we compute all runs $r_j = (\be_j, \fin_j, \per_j)$ satisfying $\be_j-1 \leq i \leq \fin_j-2\per_j$ for any $1 \leq i \leq n$.
Dumitran et al. showed how to compute these runs based on the interval stabbing algorithm~\cite{schmidt2009interval}
in total $O(n+k)$ time ($k=O(n)$ that is the number of runs).
On the other hand, we can compute such runs in our algorithm by only the operations on arrays without the complicated structures in the interval stabbing algorithm, 
since each run is checked only once in our algorithm.

We now describe this method.
Assume that all runs $r_j = (\be_j, \fin_j, \per_j)$ are sorted for $\be_j$, and for $\fin_j-2\per_j$ in increasing order.
We maintain the array $\SubRuns$ which stores runs $r_j$ satisfying $\be_j-1 \leq i \leq \fin_j-2\per_j$ for $1 \leq i \leq n$. 
We describe the state of $\SubRuns$ which stores runs satisfying $\be_j-1 \leq i \leq \fin_j-2\per_j$ as $\SubRuns_i$.
In addition, we maintain the array $P$ whose $j$-th element stores the index of run $r_j$ in $\SubRuns$. 
For each step of the computation of the $i$-th node in our algorithm ($1 \leq i \leq n$), we execute the following operations:
\begin{itemize}
  \item All runs $r_j$ satisfying $\be_j-1 \leq i \leq \fin_j-2\per_j$ are appended to $\SubRuns_i$. 
  \item We update $P[j]$ to the index in $\SubRuns_i$ of added run $r_j$. 
  \item We update $\SubRuns_i[P[j]]$ to $-1$ for all $r_j$ satisfying $\fin_j - 2\per_j = i-1$.
  \item If the value of node $i$ is $1$, we compute all runs in $\SubRuns_i$ and delete the runs. 
  After that, we delete all elements of $\SubRuns_i$.
\end{itemize}
This method is executed in $O(n)$ time, since each run is added, updated, and deleted at most once, 
and the number of runs is $O(n)$.

After all computations, if $M[n] = 1$ then any repetition factorization exists, and does not otherwise.
Thus, the following lemma clearly holds.

\begin{lemma} \label{lem:rep_1}
  Algorithm~\ref{alg:decide} can decide whether $w$ has any repetition factorizations in $O(n)$ time and space.
\end{lemma}

\begin{algorithm}[htbp]
    \begin{spacing}{0.9}
        \caption{Decide whether string $w$ has a repetition factorization}
        \label{alg:decide}
        \begin{algorithmic}[1]
          \REQUIRE string $w$ \hspace{0.5em}($|w|=n$)
          \ENSURE {\bf Yes} or {\bf No}
          \vspace{1ex}
          \STATE compute all runs $r_j(1 \leq j \leq |\runs{w}|)$
          \STATE an integer array $P$ of size $|\runs{w}|$ is initialized with $-1$
          \STATE integer array $M$ of size $n+1$ whose element $M[0]$ is initialized with $1$ and element $M[i]$ is initialized with $0$ for $1 \leq i \leq n+1$ (The index of $M$ begins with $0$.)
          \STATE integer arrays $L, R$ of size $n$ are initialized with $0$
          \FOR {\hspace{-3.2pt}{\bf each} $r_j = (\be_j, \fin_j, \per_j)$ {\bf satisfying} $\be_j = 1$}
            \STATE $L[2\per_j] \leftarrow L[2\per_j] + 1$, $R[\fin_j] \leftarrow R[\fin_j] + 1$, $P[j] \leftarrow 0$
            \STATE remove $r_j$
          \ENDFOR
          \STATE $C \leftarrow 0$
          \FOR {$i \leftarrow 1$ \KwTo $n$}
            \STATE $C \leftarrow C + L[i]$
            \IF {$C > 0$}
              \STATE $M[i] \leftarrow 1$
              \FOR {\hspace{-3.2pt}{\bf each} $r_j = (\be_j, \fin_j, \per_j)$ {\bf satisfying} $\be_j-1 \leq i \leq \fin_j-2\per_j$ ($\SubRuns_i$)}
                \STATE $L[i+2\per_j] \leftarrow L[i+2\per_j] + 1$, $R[\fin_j] \leftarrow R[\fin_j] + 1$, $P[j] \leftarrow i$
                \STATE remove $r_j$
              \ENDFOR
            \ENDIF
            \STATE $C \leftarrow C - R[i]$
          \ENDFOR
          \STATE If $M[n] = 1$ then a solution exists, and does not otherwise.
        \end{algorithmic}
        \vspace{1ex}
    \end{spacing}
\end{algorithm}

Second, we present the step of reconstructing a solution in Algorithm~\ref{alg:re}.

\begin{algorithm}[htbp]
  \begin{spacing}{0.9}
      \caption{Compute a repetition factorization of $w$}
      \label{alg:re}
      \begin{algorithmic}[1]
        \REQUIRE string $w$ \hspace{0.5em}($|w|=n$), and an integer array $P$ computed in Algorithm~\ref{alg:decide}
        \ENSURE a set $I$ of intervals for a repetition factorization of $w$
        \vspace{1ex}
        \STATE $I \leftarrow \emptyset$
        \STATE the array $F$ stores sorted runs $r_j = (\be_j, \fin_j, \per_j)$ satisfying $P[j] \neq -1$ in $w$ in descending order of $\fin_j$, and of $be_j$ if draw
        \STATE $k \leftarrow n, i \leftarrow 0$
        \WHILE {$k > 0$}
          \STATE $r_j = F[i]$
          \IF {$P[j]+ 2\per_j \leq k \leq \fin_j$}
            \STATE add an interval $[P[j]+1,k]$ to $I$
            \STATE $k = P[j]$
          \ELSE
            \STATE $i \leftarrow i+1$
          \ENDIF
        \ENDWHILE
      \end{algorithmic}
      \vspace{1ex}
  \end{spacing}
\end{algorithm}

Algorithm~\ref{alg:re} simply joins the sorted runs whose $P$ elements are not $-1$ from backward.
The sorting can be done in $O(n)$ time and space by a bucket sort.
Thus, the following lemma clearly holds.

\begin{lemma} \label{lem:rep_2}
  Algorithm~\ref{alg:re} can reconstruct a repetition factorization of $w$ in $O(n)$ time and space
  from computation of Algorithm~\ref{alg:decide}.
\end{lemma}

Theorem~\ref{th:rep} follows from Lemma~\ref{lem:rep_1} and Lemma~\ref{lem:rep_2}.

\section{Conclusions and future work}

We proposed a simple algorithm which computes an arbitrary repetition factorization in $O(n)$ time and space without relying on data structures for Union-Find and interval stabbing.
The idea of our algorithm is based on combinatorial insights of repetition factorizations, which allows us to omit redundant edges and nodes of repetition graphs.
\sinote*{modified}{%
  Also, both our algorithm and Dumitran et al's algorithm for an arbitrary repetition factorization can compute a factorization into repetitions for all prefixes $T[1..i]$ of the given string $T$, by soliving the $0$-to-$n$ reachability on the ARF-graph.
}%

Matsuoka et al.~\cite{Matsuoka16} proposed an $O(n)$-time algorithm for computing a square factorization based on the reduction
to the reachability on a similar but different run-oriented DAG.
The method of Matsuoka et al.~\cite{Matsuoka16} makes heavy use of bit operations in the word RAM model.
It is an interesting question whether we can compute a square factorization without bit operations in the word RAM model in linear time, in a similar way to our algorithm proposed in this paper.

\clearpage

\bibliographystyle{splncs04}
\bibliography{ref}

\begin{thebibliography}{10}
\providecommand{\url}[1]{\texttt{#1}}
\providecommand{\urlprefix}{URL }
\providecommand{\doi}[1]{https://doi.org/#1}

\bibitem{AlzamelISS19}
Alzamel, M., Iliopoulos, C.S., Smyth, W.F., Sung, W.: Off-line and on-line
  algorithms for closed string factorization. Theoritical Computer Science
  \textbf{792},  12--19 (2019)

\bibitem{DBLP:journals/dam/BadkobehBGIIIPS16}
Badkobeh, G., Bannai, H., Goto, K., I, T., Iliopoulos, C.S., Inenaga, S.,
  Puglisi, S.J., Sugimoto, S.: Closed factorization. Discrete Applied
  Mathematics  \textbf{212},  23--29 (2016)

\bibitem{DBLP:journals/ijfcs/BannaiGIKKPS18}
Bannai, H., Gagie, T., Inenaga, S., K{\"{a}}rkk{\"{a}}inen, J., Kempa, D.,
  Piatkowski, M., Sugimoto, S.: Diverse palindromic factorization is
  {NP}-complete. International Journal of Foundations of Computer Science
  \textbf{29}(2),  143--164 (2018)

\bibitem{RunsTheorem}
Bannai, H., I, T., Inenaga, S., Nakashima, Y., Takeda, M., Tsuruta, K.: The
  ``runs''theorem. {SIAM} Journal on Computing  \textbf{46}(5),  1501--1514
  (2017)

\bibitem{BorozdinKRS17}
Borozdin, K., Kosolobov, D., Rubinchik, M., Shur, A.M.: Palindromic length in
  linear time. In: 28th Annual Symposium on Combinatorial Pattern Matching
  ({CPM} 2017). pp. 23:1--23:12 (2017)

\bibitem{CrochemoreR95}
Crochemore, M., Rytter, W.: Squares, cubes, and time-space efficient string
  searching. Algorithmica  \textbf{13}(5),  405--425 (1995)

\bibitem{dumitran2015square}
Dumitran, M., Manea, F., Nowotka, D.: On prefix/suffix-square free words. In:
  22th International Symposium on String Processing and Information Retrieval
  (SPIRE 2015). pp. 54--66. Springer (2015)

\bibitem{Duval83}
Duval, J.: Factorizing words over an ordered alphabet. Journal of Algorithms
  \textbf{4}(4),  363--381 (1983)

\bibitem{DBLP:conf/icalp/Ellert021}
Ellert, J., Fischer, J.: Linear time runs over general ordered alphabets. In:
  48th International Colloquium on Automata, Languages and Programming ({ICALP}
  2021). LIPIcs, vol.~198, pp. 63:1--63:16 (2021)

\bibitem{DBLP:journals/jda/FiciGKK14}
Fici, G., Gagie, T., K{\"{a}}rkk{\"{a}}inen, J., Kempa, D.: A subquadratic
  algorithm for minimum palindromic factorization. Journal of Discrete
  Algorithms  \textbf{28},  41--48 (2014)

\bibitem{GABOW1985209}
Gabow, H.N., Tarjan, R.E.: A linear-time algorithm for a special case of
  disjoint set union. Journal of Computer and System Sciences  \textbf{30}(2),
  209--221 (1985)

\bibitem{DBLP:conf/cpm/ISIBT14}
I, T., Sugimoto, S., Inenaga, S., Bannai, H., Takeda, M.: Computing palindromic
  factorizations and palindromic covers on-line. In: Proc. 25th Annual
  Symposium on Combinatorial Pattern Matching (CPM 2014). pp. 150--161 (2014)

\bibitem{inoue2022factorizing}
Inoue, H., Matsuoka, Y., Nakashima, Y., Inenaga, S., Bannai, H., Takeda, M.:
  Factorizing strings into repetitions. Theory of Computing Systems
  \textbf{66}(2),  484--501 (2022)

\bibitem{KishiNI23}
Kishi, K., Nakashima, Y., Inenaga, S.: Largest repetition factorization of
  {Fibonacci} words. In: 30th String Processing and Information Retrieval
  ({SPIRE} 2023). Lecture Notes in Computer Science, vol. 14240, pp. 284--296
  (2023)

\bibitem{DBLP:conf/focs/KolpakovK99}
Kolpakov, R.M., Kucherov, G.: Finding maximal repetitions in a word in linear
  time. In: Proc. 40th Annual Symposium on Foundations of Computer Science
  (FOCS 1999). pp. 596--604 (1999)

\bibitem{Matsuoka16}
Matsuoka, Y., Inenaga, S., Bannai, H., Takeda, M., Manea, F.: Factorizing a
  string into squares in linear time. In: Proc. 27th Annual Symposium on
  Combinatorial Pattern Matching (CPM 2016). pp. 27:1--27:12 (2016)

\bibitem{schmidt2009interval}
Schmidt, J.M.: Interval stabbing problems in small integer ranges. In: 30th
  International Symposium on Algorithms and Computation (ISAAC 2009). pp.
  163--172 (2009)

\bibitem{LZ77}
Ziv, J., Lempel, A.: A universal algorithm for sequential data compression.
  IEEE Transactions on Information Theory  \textbf{IT-23}(3),  337--349 (1977)

\bibitem{LZ78}
Ziv, J., Lempel, A.: Compression of individual sequences via variable-length
  coding. IEEE Transactions on Information Theory  \textbf{24}(5),  530--536
  (1978)

\end{thebibliography}

\end{document}